\documentclass{mynature}

\usepackage{chapterbib}

\usepackage{graphics} 
\usepackage{times} 
\usepackage{mathptmx}

\usepackage[large,bf]{caption}

\typearea{14}

\onecolumn

\usepackage{setspace}

\renewcommand\refname{References}
\renewcommand{\thefootnote}{\fnsymbol{footnote}}

\usepackage{sectsty}

\begin{document}


\bibliographystyle{naturemag}

\onecolumn
\typearea{12}

\section*{\LARGE A blueprint for detecting supersymmetric dark matter
\vspace*{0.1cm}\newline in the Galactic halo}

\baselineskip20pt

\noindent{\sffamily\large  V.~Springel$^{1}$, %
S.~D.~M.~White$^{1}$, %
C.~S.~Frenk$^{2}$, %
J.~F.~Navarro$^{3,4}$, %
A.~Jenkins$^{2}$,  \vspace*{0.3cm}\\%
M.~Vogelsberger$^{1}$, %
J.~Wang$^{1}$, %
A.~Ludlow$^{3},$ %
A.~Helmi$^{5}$
\ \\

\noindent%
{\normalsize\it%
$^{1}${Max-Planck-Institute for Astrophysics, Karl-Schwarzschild-Str. 1, 85740 Garching, Germany}\\
$^{2}${Institute for Computational Cosmology, Dep.~of Physics, Univ.~of
  Durham, South Road, Durham DH1 3LE, UK}\\
$^{3}${Dep. of Physics \& Astron., University of
    Victoria, Victoria, BC, V8P 5C2, Canada}\\
$^{4}${Department of Astronomy, University of Massachusetts, Amherst,
    MA 01003-9305, USA}\\
$^{5}${Kapteyn Astronomical Institute, Univ. of Groningen,
P.O. Box 800, 9700 AV Groningen, The Netherlands}\\
}
}

\baselineskip26pt 
\setlength{\parskip}{12pt}
\setlength{\parindent}{22pt}%

\begin{large}
{\bf Dark matter is the dominant form of matter in the universe, but its nature
is unknown. It is plausibly an elementary particle, perhaps the lightest
supersymmetric partner of known particle species\cite{Bertone2005}. In this
case, annihilation of dark matter in the halo of the Milky Way should produce
$\gamma$-rays at a level which may soon be
observable\cite{Gehrels1999,Baltz2008}. Previous work has argued that the
annihilation signal will be dominated by emission from very small
clumps\cite{Calcaneo2000,Berezinsky2003} (perhaps smaller even than the Earth)
which would be most easily detected where they cluster together in the dark
matter halos of dwarf satellite galaxies\cite{Strigari2007}. Here we show, using
the largest ever simulation of the formation of a galactic halo, that such
small-scale structure will, in fact, have a negligible impact on dark matter
detectability. Rather, the dominant and likely most easily detectable signal
will be produced by diffuse dark matter in the main halo of the Milky
Way\cite{Berezinsky1994,Bergstrom1998}. If the main halo is strongly detected,
then small dark matter clumps should also be visible, but may well contain no
stars, thereby confirming a key prediction of the Cold Dark Matter (CDM) model.}
\end{large}

\begin{large}
If small-scale clumping and spatial variations in the background are neglected,
then it is easy to show that the main halo would be much more easily detected
than the halos of known satellite galaxies. For a smooth halo of given radial
profile shape, for example NFW\cite{NFW97}, the annihilation luminosity can be
written as $L\propto V_{\rm max}^4/r_{\rm half}$, where $V_{\rm max}$ is the
maximum of the circular velocity curve and $r_{\rm half}$ is the radius
containing half the annihilation flux. (For an NFW profile $r_{\rm half} =
0.089\,r_{\rm max}$, where $r_{\rm max}$ is the radius at which the circular
velocity curve peaks.)  The flux from an object at distance $d$ therefore scales
as $V_{\rm max}^4/(r_{\rm half}\, d^2)$, while the angular size of the emitting
region scales as $r_{\rm half}/d$. Hence, the signal-to-noise for detection
against a bright uniform background scales as $S/N \propto C\, V_{\rm
max}^4/(r_{\rm half}^2\, d)$.  The constant $C$ depends only weakly on profile
shape (see Supplementary Information). For the CDM simulation of the Milky Way's
halo we present below, $V_{\rm max} \simeq 209\, {\rm km\,s^{-1}}$, $r_{\rm
max}\simeq 28.4\,{\rm kpc}$ and $d\simeq 8\,{\rm kpc}$.  Using parameters for
Milky Way satellite halos from previous modelling\cite{Strigari07,Penarrubia08},
the highest $S/N$ is predicted for the Large Magellanic Cloud (LMC), for which
$V_{\rm max}\simeq 65\, {\rm km\,s^{-1}}$, $r_{\rm max}\simeq 13\,{\rm kpc}$ and
$d=48\,{\rm kpc}$, leading to $(S/N)_{\rm MW}/(S/N)_{\rm LMC}=134$!  (Note that
this overestimates the contrast achievable in practice; see Supplementary
Information.)
\end{large}

\begin{large}
The simulations used in this Letter are part of the Virgo Consortium's
Aquarius Project\cite{Springel08} to simulate the formation of CDM
halos similar to that of the Milky Way.  The largest simulation
has a dark matter particle mass of $1712~{\rm M}_\odot$ and a
converged length scale of 120~pc, both of which improve by a factor of
3 over the largest previous simulation\cite{Diemand08}. This
particular halo has mass $M_{200} = 1.84\times 10^{12}\,{\rm M}_\odot$
within $r_{200}=246\,{\rm kpc}$, the radius enclosing a mean density
200 times the critical value. Simulations of
the {\it same} object at mass resolutions lower by factors of 8,
28.68, 229.4 and 1835 enable us to check explicitly for the
convergence of the various numerical quantities presented below. 
\end{large}

\begin{large}
The detectable annihilation luminosity density at each point within a
simulation is \vspace*{0.2cm}
 \[ {\mathcal L}({\bf x}) = {\mathcal G}({\rm particle\ physics}, {\rm
  observational\ setup})\,\rho^2({\bf x}), \] where the constant ${\mathcal
  G}$ does not depend on the structure of the system but encapsulates the
properties of the dark matter particle (e.g.~annihilation cross-section and
branching ratio into photons) as well as those of the telescope and
observation. For the purposes of this Letter, we set ${\mathcal G}=1$ and give
results only for the relative luminosities and detectability of the different
structures. In this way, we can quote results that are independent of the
particle physics model and the observational details.
\end{large}

\begin{large}
Figure~1 shows the distribution of annihilation radiation within our Milky Way
halo as a function of the resolution used to simulate it. This plot excludes
the contribution to the emission from resolved substructures. Half of the
emission from the Milky Way halo is predicted to come from within $2.57\,{\rm
  kpc}$ and 95\% from within $27.3\,{\rm kpc}$.  For the lowest resolution
simulation (1835 times coarser than the largest simulation), the luminosity is
clearly depressed below 3~kpc, but for the second best simulation, it
converges well for $r >300$~pc. Thus we infer that the largest simulation
should give convergent results to $r\sim 150$~pc, and that numerical
resolution affects the luminosity of the main diffuse halo only at the few
percent level. Note that much larger effects will be caused by the baryonic
component of the Milky Way, which we neglect. This is expected to compress the
inner dark matter distribution and thus to enhance its annihilation
signal\cite{Prada2004}, which would strengthen our conclusions. (See the
Supplementary Information for discussion of this and related topics.)
\end{large}

\begin{large}

Within $433\,{\rm kpc}$ of the halo centre, we identify 297,791 and
45,024 self-bound subhalos in our two highest resolution
simulations. Many of these can be matched individually in the two
simulations, allowing a crucial (and never previously attempted) test
of the convergence of their internal structure. In Fig.~2 we show the
results of such a test. The values inferred for $V_{\rm max}$ show no
systematic offsets between simulation pairs down to the smallest
objects detected in the lower resolution simulation, suggesting that
$V_{\rm max}$ values are reliable above $\sim 1.5\,{\rm km\,s^{-1}}$
in the largest simulation.  Systematic
 offsets are visible in each simulation at small $r_{\rm max}$, reaching
 10\% on a scale 
 which decreases systematically as the resolution increases. From
 this, we conclude that our largest simulation produces $r_{\rm max}$
 values which are accurate to 10\% for $r_{\rm max} > 165\,{\rm
 pc}$. Figure~1 shows that almost all the annihilation signal from a
 halo comes from $r\ll r_{\rm max}$, corresponding to scales which are
 not well resolved for most subhalos. In the following we will
 therefore assume  the annihilation luminosity from the diffuse
 component of each subhalo to be $L= 1.23\, {\mathcal G}\,V_{\rm
 max}^4/{\rm G}^2 r_{\rm max}$, the value expected for an object with
 NFW structure.
\end{large}

\begin{large}
 When estimating the Milky Way's annihilation luminosity from our simulations,
 we need to include the following components: (1)~smooth emission associated
 with the main halo (hereafter, MainSm); (2)~smooth emission associated with
 resolved subhalos (SubSm); (3)~emission associated with unresolved substructure
 in the main halo (MainUn); and (4)~emission associated with substructure within
 the subhalos themselves (SubSub). (Here we do not discuss emission from dark
 matter caustics\cite{Mohayaee2007}.)  These 4 components have very different
 radial distributions both within the Milky Way and within its
 substructures. Neglect of this crucial fact in previous work (see below) has
 led to incorrect assessments of the importance of small-scale substructure for
 the detectability of the annihilation radiation.
\end{large}

\begin{large}
The solid blue line in Fig.~3 shows $M(<r)/M_{200}$, where $M(<r)$ is
the mass within $r$.  Half of $M_{200}$ lies within $98.5\,{\rm kpc}$
and only 3.3\% within the Solar circle ($r=8\,{\rm kpc}$). The solid
red line shows the corresponding curve for the MainSm annihilation
luminosity, normalized by $L_{200}$, its value at $r_{200}$.  This
component is much more centrally concentrated than the mass; its
half-luminosity radius is only $2.62\,{\rm kpc}$. In contrast, the
thick green line shows that the SubSm luminosity is much {\it less}
centrally concentrated than the mass. This is a result of the
dynamical disruption of substructure in the inner regions of the
halo. The thick green line includes contributions from all
substructures with mass exceeding $10^5\,{\rm M}_\odot$, almost all of
which have converged values for $V_{\rm max}$ and $r_{\rm max}$.  This
line is also normalized by $L_{200}$. Within $r_{200}$, SubSm
contributes 76\% as much luminosity as MainSm, but within $30\,{\rm
kpc}$, for example, this fraction is only 2.5\%.
\end{large}

\begin{large}
The three thin green lines in Fig.~3 show the results of excluding contributions
from less massive subhalos, corresponding to thresholds, $M_{\rm thr} = 10^6,
10^7$ and $10^8{\rm M}_\odot$. These all have similar shape and are offset
approximately equally in amplitude, implying that SubSm luminosity scales as
$M_{\rm thr}^{-0.226}$ at all radii.  If we assume, in the absence of other
information, that this behaviour continues down to a minimum mass of
$10^{-6}\,{\rm M}_\odot$, which might be appropriate if the dark matter is the
lightest supersymmetric particle\cite{Hofmann2001}, then the MainUn and SubSm
have the same radial distribution. We predict these two components together to
be 232 times more luminous than MainSm within $r_{200}$, but still only 7.8
times more luminous within $30\,{\rm kpc}$. A distant observer would thus infer
the substructure population of the Milky Way to be 232 times brighter than its
smooth dark halo, but from the Earth's position the total boost is predicted to
be only 1.9 since the substructure signal typically comes from much larger
distances.
\end{large}

\begin{large}
We must now consider the additional luminosity due to
(sub-)substructures (SubSub). Before a subhalo is accreted onto the
main object, we assume its detailed structure to be similar to that of
the main halo (including its subhalo population), but scaled down
appropriately in mass and radius. (We have checked that such a scaling
does indeed hold approximately for small independent objects outside
the main halo.)  However, once the subhalo is accreted, its outer
regions are rapidly removed by tidal stripping. The longer a subhalo
has been part of the main system and the closer it is to the centre,
the more drastic is the stripping\cite{DeLucia2004,Gao04}. As a
result, most of the substructure associated with the subhalo is
removed, while its smooth luminosity is little affected. The removed
(sub-)subhalos are, in effect, transferred to components SubSm and
MainUn.
\end{large}

\begin{large}
A subhalo at Galactocentric distance $r$ is typically truncated at
tidal radius $r_t= (M_{\rm sub}/[(2-{\rm dln}M/{\rm
dln}r)\,M(<r)])^{1/3}r$. We estimate its SubSub luminosity by assuming
that all material beyond $r_t$ is simply removed. The remaining SubSub
luminosity can then be obtained from the curves of Fig.~3 if we scale
them to match the measured parameters of the subhalo ($M_{\rm sub}$,
$V_{\rm max}$ and $r_{\rm max}$).  We assume that the $r_{200}$ of the
subhalo before accretion was proportional to its present $V_{\rm
max}$. ($r_{200}$ is indeed nearly proportional to $V_{\rm max}$ for
isolated halos in our simulations.)  We further assume that the ratio
of subhalo mass to SubSub luminosity within $r_t$ corresponds to the
ratio between main halo mass and SubSm luminosity (from Fig.~3) within
the scaled radius $r_t/f$, where $f=(V_{\rm max}/209\,{\rm
km\,s^{-1}})$. We must also correct for the SubSub luminosity below
the mass limit $M_{\rm min}=10^5 f^3~{\rm M}_\odot$, scaling down the
resolution limit of our simulation appropriately for the subhalo. The
SubSub luminosity must then be boosted by a factor $(M_{\rm min}/
M_{\rm lim})^{0.226}$ where $M_{\rm lim}$ is the free-streaming mass,
$10^{-6}\,{\rm M}_\odot$, in the example given above. For
definiteness, we adopt $M_{\rm lim} = 10^{-6}\,{\rm M}_\odot$ in the
discussion below, although none of our conclusions would change if we
adopted, for example, $M_{\rm lim} = 10^{-12}\,{\rm M}_\odot$.
\end{large}

\begin{large}
We now consider the expected appearance and detectability of these various
components. The diffuse emission from the Milky Way's halo (MainSm) is
distributed across the whole sky falling away smoothly from the Galactic
centre. A randomly placed observer at $r=8\,{\rm kpc}$ sees half the flux
within 13~degrees of the Galactic centre, most of this well outside the
Galactic plane where contamination is strongest. Assuming NFW structure for
individual subhalos, half the diffuse emission from each object falls within
the angular radius corresponding to $r_{\rm half}= 0.089\,r_{\rm
  max}$. Because of their large typical distances, these subhalos are almost
uniformly distributed across the sky. The luminosity from unresolved subhalos
(MainUn) is similarly distributed and will appear smooth in $\gamma$-ray sky
maps, with a centre to anticentre surface brightness contrast of only
1.54. Half the luminosity from (sub-)subhalos within an individual subhalo
falls within an angular radius corresponding to $\sim 0.6\,r_t$; this is
usually much more extended than the SubSm emission from the same subhalo.
\end{large}

\begin{large}
This information allows us finally to calculate the relative
detectability of the various components. As argued above, the
signal-to-noise for detection by an optimal filter against a bright
uniform background can be written as $S/N\propto F/\theta_h$, where
$F$ is the total flux, $\theta_h$ is the angle containing half this
flux and the constant of proportionality depends weakly on profile
shape but strongly on the particle-physics and observational
parameters (the factor ${\mathcal G}$ above). To account for the
finite angular resolution of the observation, we replace $\theta_h$
with $\theta_h' = (\theta_h^2 + \theta_{\rm psf}^2)^{1/2}$. For
example, $\theta_{\rm psf} \simeq 10\,{\rm arcmin}$ is the
characteristic angular resolution of the LAT detector of the recently
launched GLAST telescope at the relevant
energies\cite{Michelson2007}. In reality, the background at these
energies is not uniform and is relatively poorly
known\cite{Hunter1997,Strong2004}. In the Supplementary Information,
we show that this is likely to reduce the detectability of the main
smooth halo relative to that of subhalos by a factor up to ten
compared to the numbers we quote below which are based on the simple
assumption of a uniform background.
\end{large}

\begin{large}
In Fig.~4 we combine data for 1000 randomly placed observers at
$r=8\,{\rm kpc}$. Panel (a) shows histograms of the $S/N$ for
detecting SubSm and SubSub emission from the 30 highest $S/N$
subhalos, and also shows the expected $S/N$ for known satellites of
the Milky Way. These are all expressed in units of the $S/N$ for
detecting the MainSm emission. Three important conclusions follow
immediately: (1)~no subhalo is expected to have $S/N$ more than $\sim
10$\% that of the main halo even accounting for the expected effects
of the non-uniform background; (2)~the most easily detectable dark
subhalo is predicted to have 5 times larger $S/N$ than the LMC; and
(3)~the $S/N$ predicted for SubSub emission is always much lower than
that predicted for SubSm emission because of the much greater angular
extent of the former.
\end{large}

\begin{large}
Panels~(b) to~(f) of Fig.~4 show histograms of the masses, $V_{\rm
max}$ values, distances, angular half-light radii, and fluxes
(relative to the flux from the main halo) of the 30 highest $S/N$
subhalos. These are compared with the distributions for the known
satellites of the Milky Way where appropriate. For the fluxes and
half-light radii we show separate histograms for the SubSm and SubSub
emission. A second set of important conclusions follow. If subhalos
are detected, then the highest $S/N$ systems: (4)~will typically have
masses and circular velocities well below those inferred for the
currently known satellites of the Milky Way; (5)~will have angular
half-light radii below 10 arcmin and so will not be resolved by GLAST;
(6)~will be at distances $\sim 4$ kpc; and (7)~will typically have
SubSm and SubSub fluxes which are factors of $10^{-4}$ and $10^{-6}$
times lower than those of the main halo, respectively.
\end{large}

\begin{large}
These conclusions differ substantially from earlier work. Very small-scale
substructure (below the resolution limit of our simulations) does not affect the
detectability of dark matter annihilation in the Milky Way's halo. This is true
both for the smooth main halo (contradicting references 4, 5 and 22
amongst others) and for its subhalos (contradicting 6, 23 and 24).
Emission should be much more easily detected from the main halo than from
subhalos (contradicting 25 and 26, 
but in agreement with 27),
even though the total flux is dominated by substructures (contradicting
28 and 29).
The most easily detectable subhalo is expected to be a relatively nearby object
of lower mass than any known Milky Way satellite (contradicting 23 and 25).
Almost all of these differences stem from the differing spatial distribution of
small-scale substructure and smooth dark matter which our simulations are able
to trace reliably because of their high resolution.
\end{large}

\begin{large}
The GLAST satellite is now in orbit and accumulating a $\gamma$-ray
image of the whole sky. If Nature obeys supersymmetry and the
parameters of the theory are favourable, in a few years we may have a
direct image of the Galaxy's dark halo. If we are really lucky, we may
also detect substructures both without and with stars. This would
provide a convincing confirmation of the Cold Dark Matter theory.
\end{large}

\begin{large}
\bibliography{paper}
\end{large}

\paragraphfont{\large}

\begin{large}
\vspace*{-0.5cm}\paragraph*{Competing interests} The authors declare that they have no
competing financial interests.
\end{large}

\begin{large}
\vspace*{-0.5cm}\paragraph*{Correspondence} and requests for materials should
be addressed to V.S.~(email: vspringel@mpa-garching.mpg.de).
\end{large}

\newpage

\begin{figure}
\begin{center}
\resizebox{16.0cm}{!}{\includegraphics{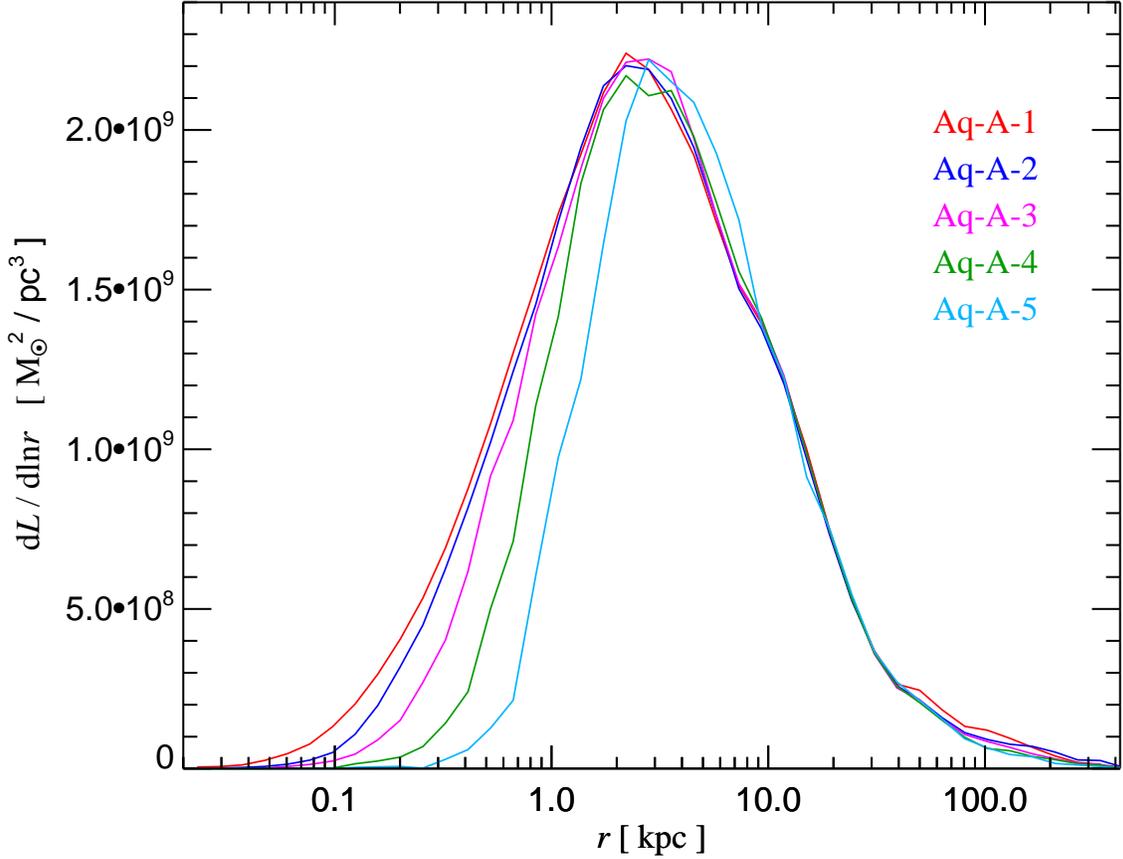}}%
\caption{  \baselineskip22pt
    Annihilation luminosity as a function of radius for the diffuse dark
    matter component of Milky Way halos. These simulations assume
    ${\mathcal G}=1$ and a Universe with mean matter density
    $\Omega_{\rm m}=0.25$, cosmological constant density
    $\Omega_{\Lambda}=0.75$, Hubble constant $H_0=73\,{\rm km\,
    s^{-1}Mpc^{-1}}$, primordial spectral index $n=1$ and present
    fluctuation amplitude $\sigma_8=0.9$.  In this representation, the
    total
    emitted luminosity is proportional to the area under each
    curve. The particle mass in the simulations is $1712\, {\rm
    M}_\odot$ for Aq-A-1, and grows to $1.37\times 10^4$,
    $4.91\times 10^4$, $3.93\times 10^5\,{\rm M}_{\odot}$ and
    $3.14\times 10^6\,{\rm M}_{\odot}$ for  Aq-A-2, Aq-A-3, Aq-A-4
    and Aq-A-5, respectively. The fluctuations at
    large radii are due to subhalos below our 
    detection limit. These curves were calculated by estimating a
    density local to each N-body particle through a Voronoi
    tesselation of the full particle distribution and then summing the
    annihilation luminosities of individual particles in a set of
    logarithmically spaced spherical shells. Note that the vertical
    axis is linear, so these curves demonstrate numerical convergence
    at the percent level in the detailed structure of our main halo
    down to scales below $1\,{\rm kpc}$. 
    \label{FigA}}
\end{center}
\end{figure}

\newpage

\begin{figure}
\begin{center}
\vspace*{-1.5cm}\resizebox{11.5cm}{!}{\includegraphics{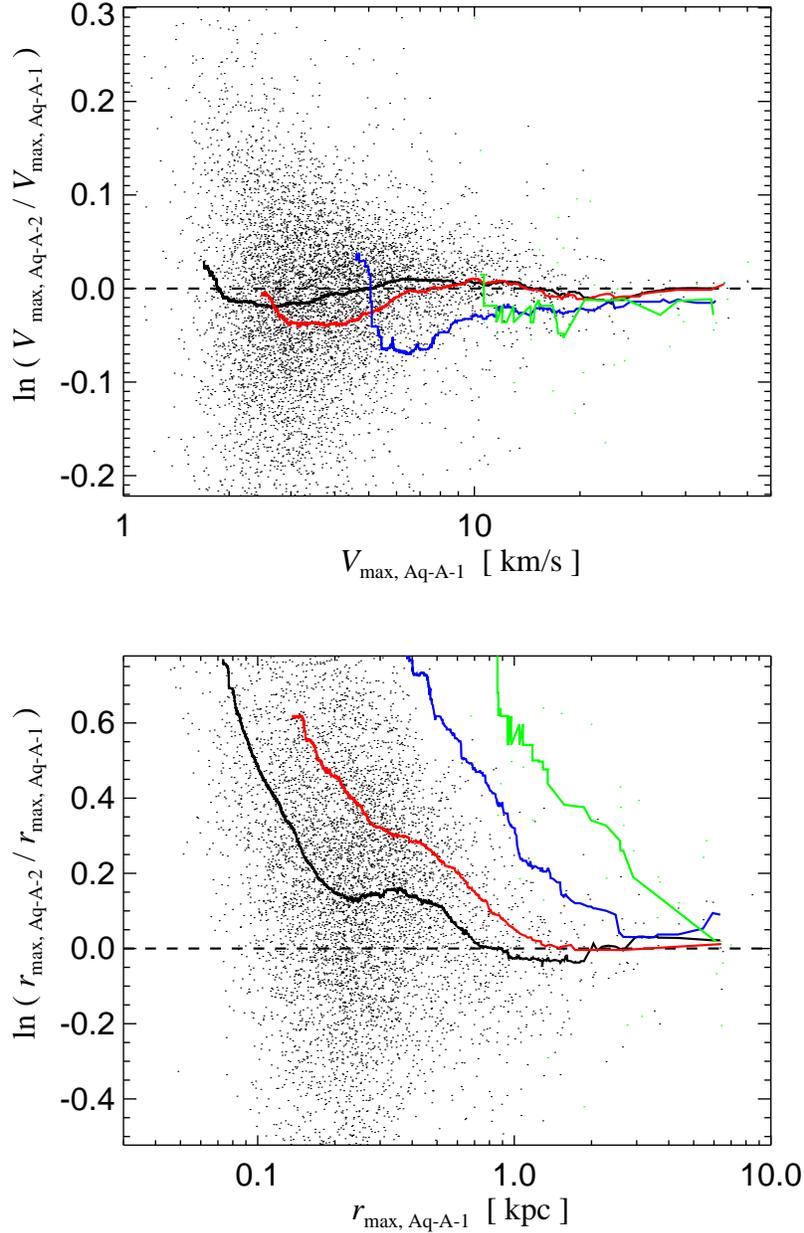}}%
\caption{ \baselineskip22pt Structural properties of dark matter subhalos as a function of
  simulation resolution. {\it Upper panel}: ${\rm ln} (V_{{\rm max,\,
      Aq-A-2}}/V_{\rm max,\, Aq-A-1})$ against $V_{{\rm max,\, Aq-A-1}}$ for
  6711 matched subhalos detected by the {\small SUBFIND}
  algorithm\cite{Springel2001} within $433\,{\rm kpc}$ of halo centre in our
  two highest resolution simulations, Aq-A-1 and Aq-A-2. The radius $433\,{\rm
    kpc}$ encloses an overdensity 200 times the cosmic mean. The black solid
  line shows the running median of this distribution. Red, blue and green
  lines give similar median curves for matches of the lower resolution
  simulations to the highest resolution simulation, Aq-A-1. {\it Lower panel}:
  as above but for the ratio of characteristic sizes ($r_{\rm max}$) as a
  function of that in the highest resolution simulation. We have checked that
  convergence in the subhalo mass is similarly good and that these results
  apply equally well to subhalos inside $50\,{\rm kpc}$.}
\end{center}
\end{figure}

\newpage

\begin{figure}
\begin{center}
\resizebox{16.0cm}{!}{\includegraphics{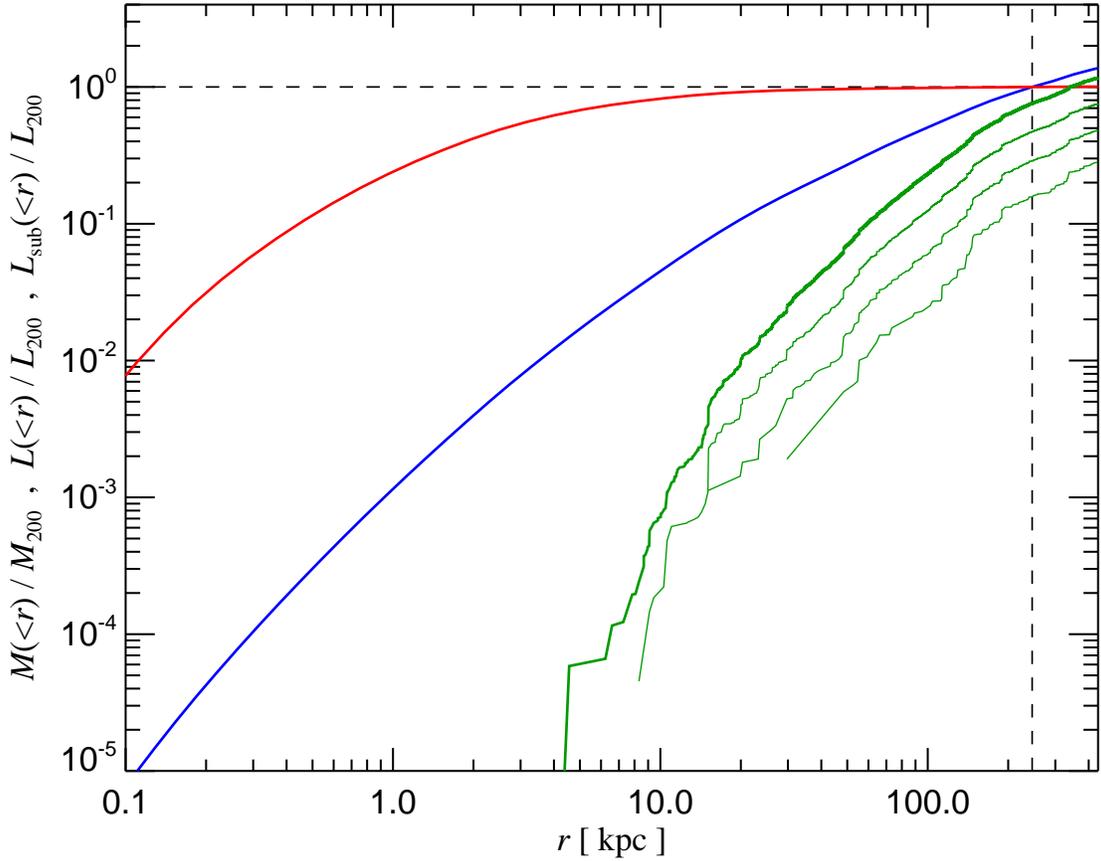}}%
\caption{ \baselineskip22pt Radial dependence of the enclosed mass and annihilation
luminosity of various halo components. The blue line gives enclosed
dark matter mass in units of $M_{200}$, the value at $r_{200}$ (the
radius enclosing a mean density 200 times the critical value, 
marked in the plot by a vertical dashed line). The red line gives the
luminosity of smooth main halo annihilation (MainSm) in units of
$L_{200}$, its value at $r_{200}$. The green lines give the luminosity
of smooth subhalo annihilation (SubSm) for various lower limits
to the subhalo mass considered; the solid thick line is for $M_{\rm
min} = 10^5\,{\rm M}_\odot$, the thin lines for $M_{\rm min} = 10^6,
10^7$ and $10^8\,{\rm M}_\odot$. Note that the shape of these lines is
insensitive to $M_{\rm min}$, and that their normalization is
proportional to $M_{\rm min}^{-0.226}$.}
\end{center}
\end{figure}
\clearpage

\newpage

\begin{center}
\resizebox{16.0cm}{!}{\includegraphics{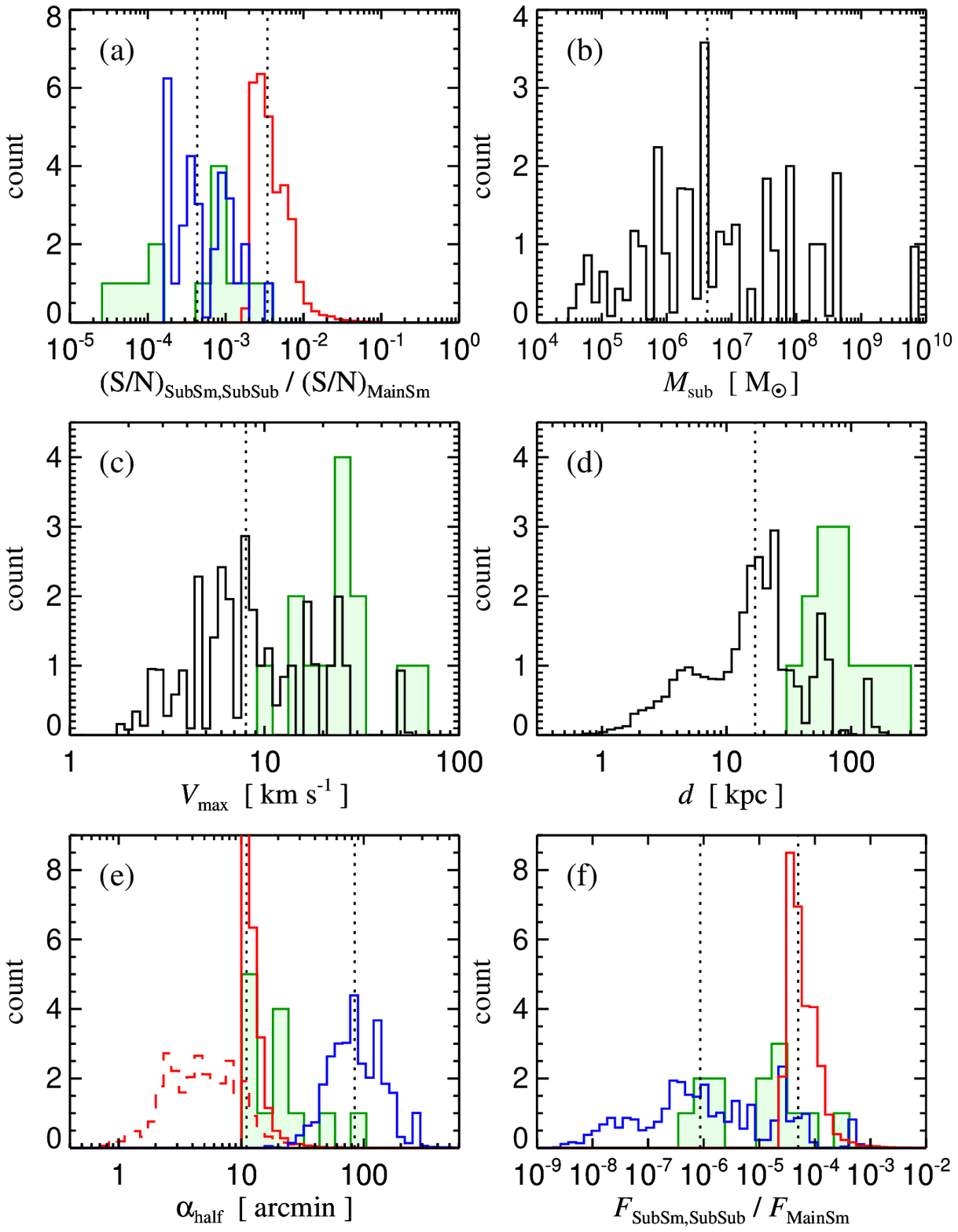}}
\end{center}

\begin{figure}
\begin{center}
  \caption{ \baselineskip22pt Observability of subhalos. The histograms show the properties of
    the 30 highest $S/N$ subhalos seen by each of 1000 observers placed at
    random 8 kpc from halo centre, assuming a 10 arcmin observational
    beam. The histograms are divided by 1000 so they sum to 30. Panel~(a): the
    30 highest $S/N$ values for SubSm (red) and SubSub (blue) emission. These
    do not necessarily come from the same subhalos. The SubSub $S/N$ values
    lie well below the SubSm values -- subhalo detectability is not influenced
    by internal substructure. Panels~(b), (c) and~(d): histograms of the
    masses, $V_{\rm max}$ values and distances of the 30 halos with highest
    $S/N$ SubSm emission. For these same halos panels~(e) and~(f) show
    half-light radii and fluxes separately for the SubSm (red) and the SubSub
    (blue) emission. In panel~(e) the dashed and solid red histograms show
    values before and after convolution with the telescope beam.  For subhalos
    with $V_{\rm max} < 5\,{\rm km\,s^{-1}}$ we have suppressed numerical
    noise by replacing the measured $r_{\rm max}$ by a value drawn from a
    suitably scaled version of the distribution measured at larger $V_{\rm
      max}$ for subhalos within 50~kpc. This substitution has a modest effect
    on the low mass tails of our distributions. Fluxes are expressed in units
    of the flux from the main halo. Dashed vertical lines mark median
    values. The single highest $S/N$ subhalos detected by each of our 1000
    ``observers'' are biased towards smaller and nearer objects; their median
    values are $(S/N)_{\rm SubSm}= 0.015\,(S/N)_{\rm MainSm}$, $V_{\rm
      max}=6\,{\rm km\,s^{-1}}$, $M_{\rm sub} = 2\times 10^6\, {\rm M}_\odot$
    and $d=4\,{\rm kpc}$. Light green histograms show the distributions
    predicted for SubSm emission from 13 known satellites of the Milky Way,
    based on published mass models\cite{Penarrubia08}.}
\end{center}
\end{figure}


\paragraphfont{\small}
\subsectionfont{\normalsize}

\def\captionfont{\small}

\def\@cite#1#2{[$^{\mbox{\scriptsize #1\if@tempswa , #2\fi}}$]}

\typearea{14}

\renewcommand\refname{References}
\renewcommand{\thefootnote}{\fnsymbol{footnote}}

\newcommand\be{\begin{equation}}
\newcommand\ee{\end{equation}}

\renewcommand{\textfraction}{0}
\renewcommand{\floatpagefraction}{1.0}
\renewcommand{\topfraction}{1.0}
\renewcommand{\bottomfraction}{1.0}

\title{\vspace*{0.3cm} \ \\ {\Large \em Supplementary Information} \vspace*{0.3cm}}


\author{\parbox{9.5cm}{
{\small\sffamily%
V.~Springel$^{1}$, %
S.~D.~M.~White$^{1}$, %
C.~S.~Frenk$^{2}$, %
J.~F.~Navarro$^{3,4}$, %
A.~Jenkins$^{2}$, %
M.~Vogelsberger$^{1}$, %
J.~Wang$^{1}$, %
A.~Ludlow$^{3},$ %
A.~Helmi$^{5}$}}}

\date{}

\bibliographystyle{naturemag}

\baselineskip14pt 
\setlength{\parskip}{4pt}
\setlength{\parindent}{18pt}%

\twocolumn

\setlength{\footskip}{25pt}
\setlength{\textheight}{640pt}


\maketitle

\renewcommand{\thefootnote}{\arabic{footnote}}
\footnotetext[1]{Max-Planck-Institute for Astrophysics, Karl-Schwarzschild-Str.~1, 85740 Garching, Germany}
\footnotetext[2]{Institute for Computational Cosmology, Dep. of
    Physics, Univ. of Durham, South Road, Durham  DH1 3LE, UK}
\footnotetext[3]{Dep. of Physics \& Astron., University of Victoria, Victoria,
  BC, V8P 5C2, Canada}
\footnotetext[4]{Dep. of Astron., University of Massachusetts, Amherst, MA 01003-9305, USA}
\footnotetext[5]{Kapteyn Astronomical Institute, Univ. of Groningen,
P.O. Box 800, 9700 AV Groningen, The Netherlands}

{\bf This document provides supplementary information for our Letter to
  Nature. In particular, we detail the signal-to-noise calculation and the
  treatment of the background.}  \vspace*{0.15cm}

\renewcommand{\thefootnote}{\fnsymbol{footnote}}

\small 

\subsection*{Optimal filter for detection}

The search for $\gamma$-ray photons from dark matter annihilation is made
difficult by the low photon flux received here on Earth (or in orbit for that
matter), which introduces the familiar limitations of Poisson statistics. In
this case, the discrimination against the background flux and other sources of
noise can benefit from an optimal filter for detection.

For definiteness, let $n_\gamma(\theta, \phi)$ describe the mean specific intensity
of~$\gamma$-ray photons from dark matter annihilation on the sky (i.e. photons
per second, per unit area, and unit solid angle). Further, let $b_\gamma(\theta,
\phi)$ be the specific intensity of background photons. We define as
expected signal
the quantity
\begin{equation}
\left<S\right> =\tau\, A_{\rm eff}  \,\int w(\theta, \phi)\,n_\gamma(\theta, \phi) {\rm d}\Omega ,
\end{equation}
where $w(\theta,\phi)$ is a suitable {\em filter} function, $\tau$ is the
observational integration time, and $A_{\rm eff}$ is the effective area of the
telescope. Likewise, the total expected photon count within the same filter function is
\begin{equation}
 \left< T \right> = \tau\, A_{\rm eff}  \int w(\theta, \phi)\,[n_\gamma(\theta, \phi) +b_\gamma(\theta, \phi)]  {\rm d}\Omega .
\end{equation}
To establish the significance of the signal, we need to compare its
expectation value to the dispersion of the total photon count. We define
as {\em signal-to-noise} ratio the quantity
\begin{equation}
(S/N) = \left<S\right> / \left[ \left<T^2\right> - \left<T\right>^2 \right]^{1/2}.
\end{equation}
Since the arriving photons sample the signal and the background distribution
as a Poisson process, we have
\begin{equation}
(S/N)^2 = \tau\, A_{\rm eff}\frac{ [\int w(\theta, \phi)\,n_\gamma(\theta, \phi) {\rm d}\Omega ]^2}
{\int w^2(\theta, \phi)\,[n_\gamma(\theta, \phi) +b_\gamma(\theta, \phi)]  {\rm d}\Omega }.
\end{equation}
It is easy to show that the signal-to-noise is maximized if the filter $w$ is
chosen as
\begin{equation}
w(\theta, \phi) \propto \frac{n_\gamma(\theta, \phi)}{n_\gamma(\theta, \phi) +b_\gamma(\theta, \phi)}.
\end{equation}
For this {\em optimal filter}, the resulting signal-to-noise is then given by
\begin{equation}
  S/N = \sqrt{\tau\, A_{\rm eff}} \left[ \int \frac{ n_\gamma^2(\theta, \phi)}{ n_\gamma(\theta, \phi) + b_\gamma(\theta, \phi)} {\rm d}\Omega \right]^{1/2}.
\label{EqSN}
\end{equation}

Note that {\em if the background dominates} (the regime relevant for dark
matter annihilation in the Milky Way, except perhaps in the very centers of
halos and subhalos), $n_\gamma$ can be neglected against $b_\gamma$ in the
denominator above, and the optimal filter shape is simply proportional to the
signal profile divided by the background profile. If the variation of the
background over the source can be neglected, the optimal filter is just given
by the signal shape. Another important consequence of background dominance is
that the overall amplitude of the background drops out when the
signal-to-noise of two different dark matter structures in the Milky Way is
compared.

\subsection*{Signal-to-noise calculation for different flux components}

We now calculate the optimal $S/N$ for the different components in our
Milky Way simulation model. To account for the finite angular
resolution of the telescope (of order $\theta_{\rm psf} \sim 10\,{\rm
  arcmin}$ for GLAST at the relevant energies), we convolve the raw
infinite-resolution signal $n_\gamma$ with a Gaussian
point-spread-function (PSF),
\begin{equation}
w_{\rm psf}(\vec{\alpha}) = \frac{\ln 2}{\pi\theta_{\rm
    psf}^2}\exp\left(-\ln 2 \frac{\vec{\alpha}^2}{\theta_{\rm psf}^2}\right) ,
\end{equation}
before using it to calculate the $S/N$ with equation (\ref{EqSN}). We
have here chosen to parameterize the PSF with the angle $\theta_{\rm
  psf}$ that contains half the light, for consistency with the
half-light angles we use to characterize the sources. (The angle that
contains 68\% of the light is larger than $\theta_{\rm psf}$ by a
factor 1.28.)  Note that smoothing allows us to continue to use the
assumption of background dominance even for cuspy dark matter halos,
which in the infinite resolution case can have a diverging surface
brightness in their centre.

\subsubsection*{Component ``SubSm'',  resolved subhalos}

\begin{figure} 
\begin{center} 
\resizebox{8cm}{!}{\includegraphics{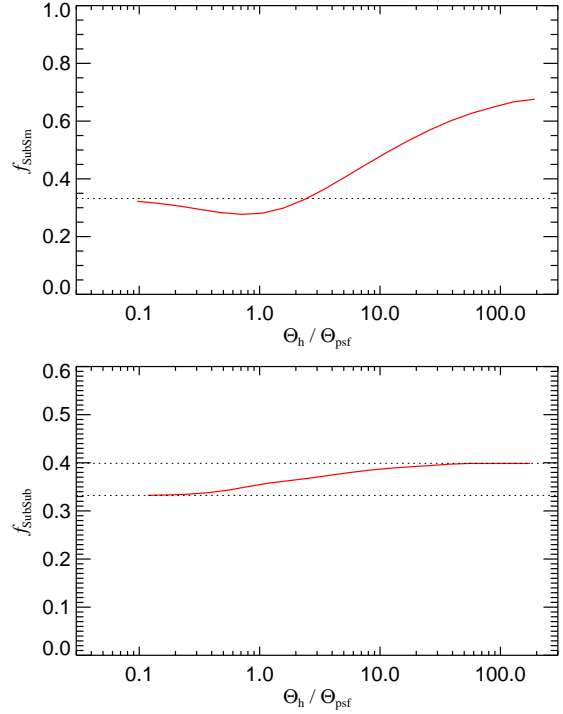}}
\end{center}
\caption{Signal-to-noise prefactors as a function of angular resolution.
The top panel gives the prefactor 
for a subhalo with  NFW profile seen in projection, while the bottom panel
gives the prefactor for its much more extended (sub-)substructure emission.
\label{FigSignalNoisePrefactors}}
\end{figure}

We assume that the background does not vary significantly over the angular
extent of a subhalo at location $\vec{\alpha}$. It is then numerically
straightforward first to  calculate the specific intensity of a given subhalo
on the sky (characterized by its total luminosity, distance, and half-light
angle), then to convolve it with the Gaussian PSF, and finally to calculate
its $S/N$ with equation (\ref{EqSN}). The result can be written as
\begin{equation}
  (S/N)_{\rm SubSm} =  f_{\rm SubSm} \left(\theta_h/\theta_{\rm psf}\right) 
  \left[ \frac{\tau\, A_{\rm eff} }{b_\gamma(\vec{\alpha})}\right]^{1/2} \frac{F}{(\theta_h^2 +
    \theta_{\rm psf}^2)^{1/2}} \, ,
\end{equation}
where $F=L/(4\pi d^2)$ is the photon flux of the source, and \linebreak $f_{\rm
  SubSm}(x)$ is a factor of order unity which depends on the detailed shape of
the source. As the shape varies depending on  whether the source is
resolved or not, $f_{\rm SubSm}$ is a function of the ratio
$x=\theta_h/\theta_{\rm psf}$ of the half-light angle and the angular
resolution. In the top panel of Figure~\ref{FigSignalNoisePrefactors}, we show
$f_{\rm SubSm}$ for projected NFW halos, our adopted model for the SubSm
emission. For a source that is completely unresolved, $\theta_h \ll
\theta_{\rm psf}$, the factor $f_{\rm SubSm}$ approaches $\sqrt{(\ln 2)/2
  \pi}\simeq 0.332$, the value expected for a Gaussian signal. Note however
that the overall variation of $f_{\rm SubSm}$ between unresolved and well
resolved subhalos is very modest, and at most a factor $\sim 2$.

\subsubsection*{Component ``SubSub'', substructures in subhalos}

As we have shown in this Letter, the radial distribution of substructure is
strongly antibiased with respect to the mass of the parent halo. This in turn
applies to \mbox{(sub-)substructure} within a subhalo. An immediate
consequence is that the surface brightness profile of (sub-)substructure is
much more extended than that of the parent subhalo, and unlike that of the
parent halo it is strongly affected by tidal truncation and the associated
mass loss in the outer parts of the halo.

\begin{figure} 
\begin{center} 
\resizebox{8.2cm}{!}{\includegraphics{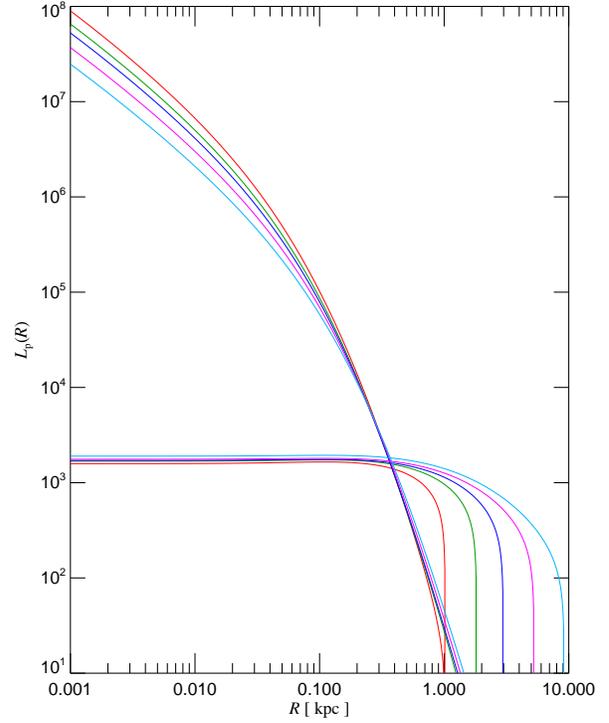}}
\end{center}
\caption{Comparison of the shape of the surface brightness profile of
  typical subhalos with $V_{\rm max}=10\,{\rm km\,s^{-1}}$ at
  different Galacto-centric distances equal to $d=400$, $200$, $50$ and $12.5\,{\rm
    kpc}$ from the halo center, as seen by a distant observer. The
  upper set of solid lines shows the surface brightness profile of the
  subhalo's diffuse emission (component SubSm), which we model as a
  NFW profile that is truncated at the tidal truncation radius. The
  lower set of solid lines gives the surface brightness profile of all
  \mbox{(sub-)substructure} emission (component SubSub) in the subhalo,
  extrapolated down to $10^{-6}\,{\rm M}_\odot$, while the dotted
  lines show the total profile. Note that the central surface
  brightness of the smooth emission as well as its total luminosity
  actually increase nearer to the centre of the main halo, because subhalos of equal
  $V_{\rm max}$ are on average more concentrated at smaller distances. On the other hand, the
  \mbox{(sub-)substructure} emission has a very different, nearly flat
  surface brightness profile, because most substructures are located
  in the outer parts of the subhalo. As a result, tidal truncation
  strongly reduces the luminosity of this component and leads to a
  decline of the SubSub surface brightness for smaller distances.
  \label{FigCompSurfBrightnessProfile}}
\end{figure}

We illustrate this qualitatively very different behaviour in
Figure~\ref{FigCompSurfBrightnessProfile}, which shows the azimuthally
averaged surface brightness profile we adopt for typical $V_{\rm
  max}=10\,{\rm km\,s^{-1}}$ subhalos at various distances from the
Galactic centre and so with different tidal radii $r_t$. Each profile
is split into the contribution from smooth DM (SubSm) and from
\mbox{(sub-)substructures} (SubSub).  Interestingly, the substructure surface
brightness is nearly constant in projection. This remains the case
when the halo is sharply tidally truncated at progressively smaller
radii. The tidal truncation results in the loss of substructures so that the
total SubSub luminosity declines strongly, yet the shape of the
luminosity profile remains remarkably flat and can be approximated as
a disk on the sky that becomes progressively fainter and smaller.  In
contrast, the total SubSm luminosity is little affected by tidal
truncation and the main systematic effect with Galacto-centric
distance is that subhalos of given mass become more concentrated with
decreasing Galacto-centric distance, thus enhancing their SubSm
luminosity.

For the purposes of calculating the signal-to-noise of the SubSub component,
we assume that its surface brightness profile can be modelled as a uniform
disk, subject to the smoothing of the finite angular resolution of the
telescope.  Again, we can write the resulting signal-to-noise in the generic
form
\begin{equation}
(S/N)_{\rm SubSub} =  f_{\rm SubSub} \left(\theta_h/\theta_{\rm psf}\right) 
\left[ \frac{\tau\, A_{\rm eff} }{b_\gamma(\vec{\alpha})}\right]^{1/2} \frac{F}{(\theta_h^2 +
  \theta_{\rm psf}^2)^{1/2}} \, ,
\end{equation}
where $F=L/(4\pi d^2)$ is the photon flux of the component, and $\theta_h$ is
its half-light radius.  In the bottom panel of
Figure~\ref{FigSignalNoisePrefactors}, we show the variation of $f_{\rm
  SubSub}$ with angular resolution. The factor $f_{\rm SubSub}$ varies smoothly between
the value $\sqrt{(\ln 2)/2 \pi}\simeq 0.332$ expected for an unresolved
source, to the value $ 1/\sqrt{2 \pi}\simeq 0.399$ expected for a well
resolved disk-like component.

\subsubsection*{Component ``MainSm'', smooth main halo}

The main halo's half-light angle of 13 degrees for the diffuse emission is
resolved well with GLAST, but as our simulated emission maps have in principle
higher resolution, we also smooth them with the PSF before calculating the
$S/N$, for consistency. For definiteness we also average the main halo's
diffuse emission in angular rings around the Galactic centre, and for 25
random observer positions. We note however that the dispersion arising from
different observer viewpoints and orientations of the galactic plane assumed
within the simulated halo is quite small.

A problematic point for the detection of the diffuse emission is the strong
and variable background from other Galactic $\gamma$-ray emission, resulting
mostly from the interaction of cosmic rays with the interstellar gas and from
inverse Compton upscattering by electrons.  A considerable body of literature
is concerned with modelling the background and explaining the spectra observed
with EGRET, in particular. The best predictions for the Galactic diffuse
$\gamma$-ray background are obtained with complex  codes like
GALPROP [31], which treat CR interactions and propagation in the
Galaxy in detail. However, the modelling uncertainties are still substantial,
so we consider a few limiting cases for the spatial variation of the
background.

\begin{figure} 
\begin{center} 
\resizebox{8.2cm}{!}{\rotatebox{90}{\includegraphics{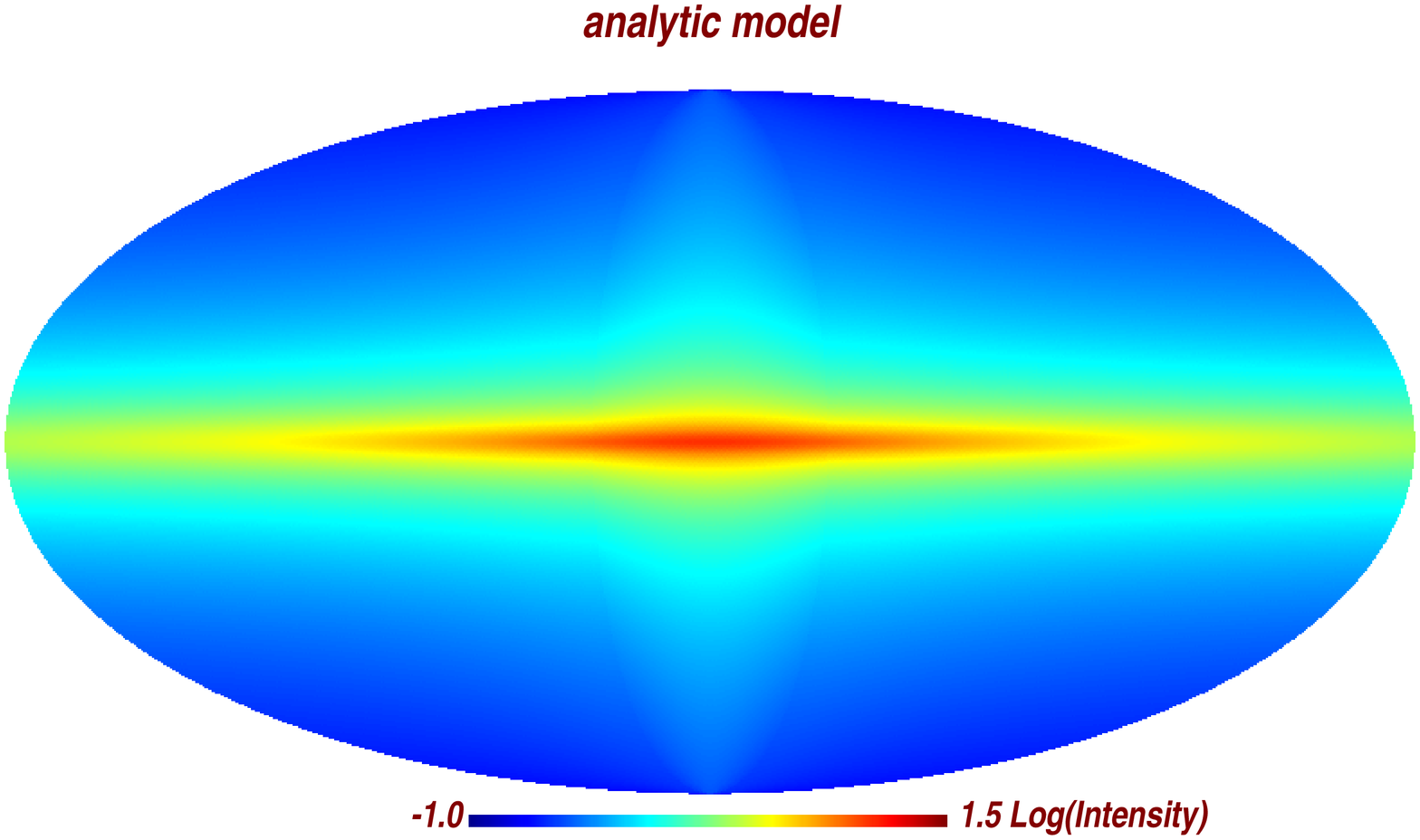}}}\\
\ \\
\resizebox{8.2cm}{!}{\rotatebox{90}{\includegraphics{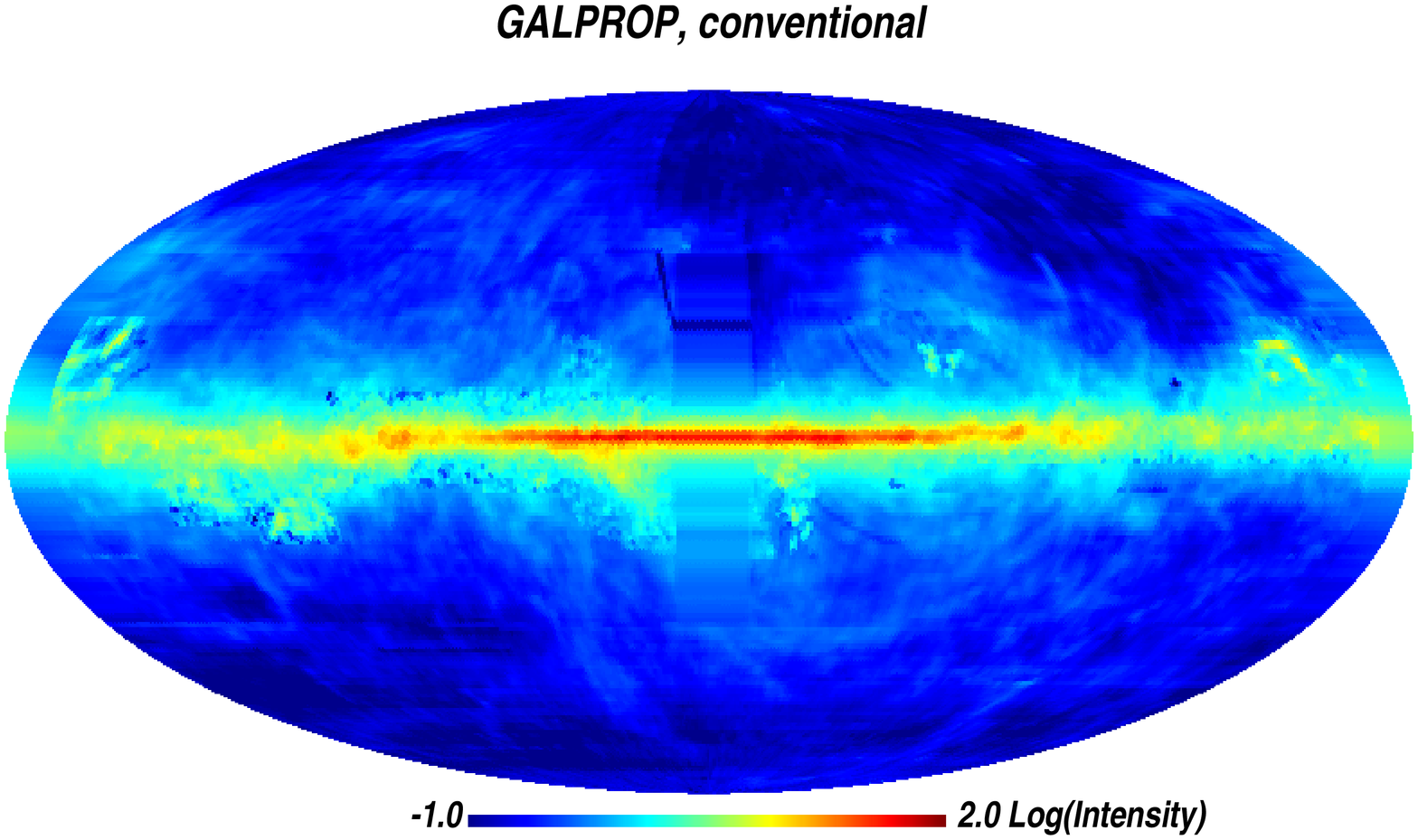}}}\\
\ \\
\resizebox{8.2cm}{!}{\rotatebox{90}{\includegraphics{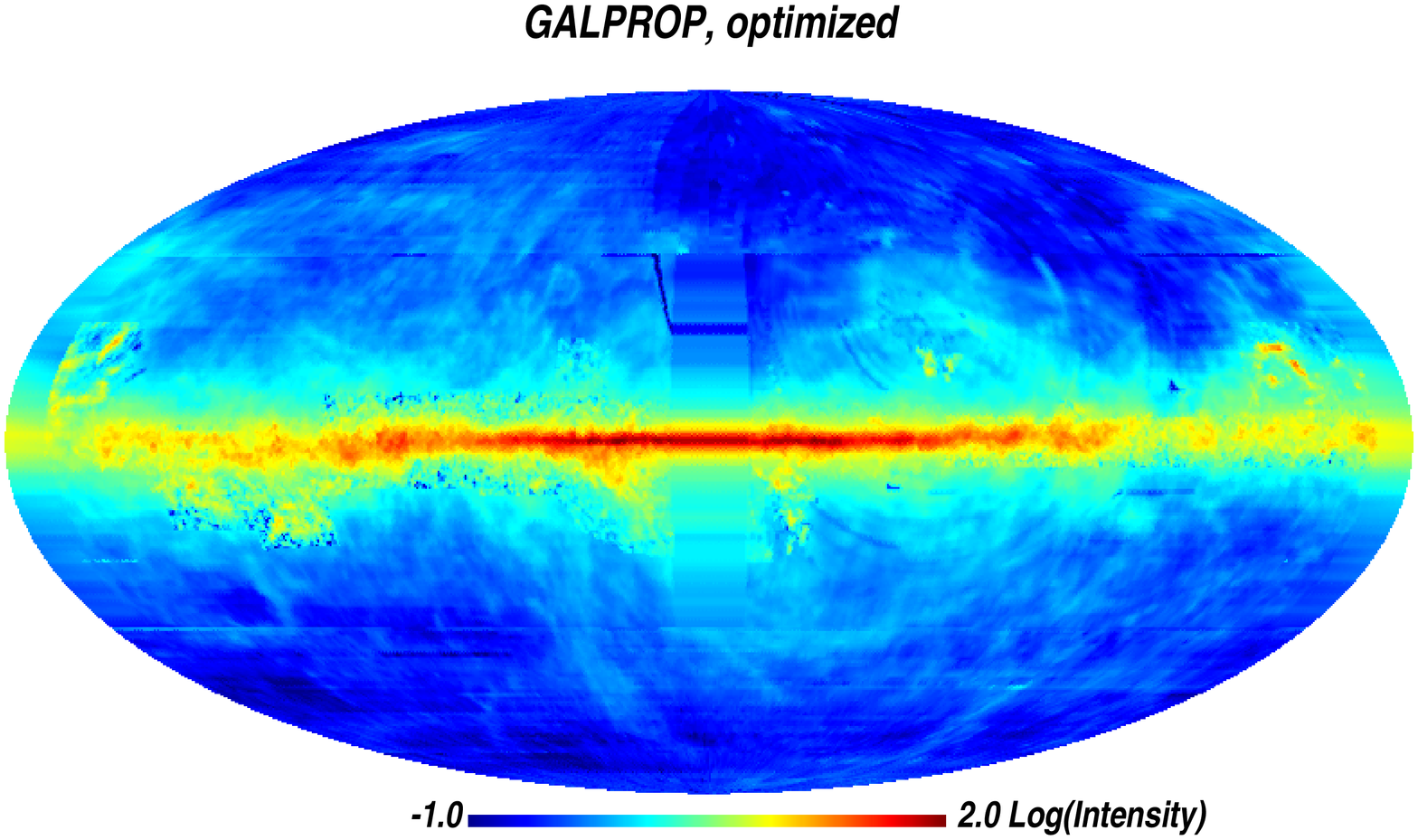}}}
\end{center}
\caption{Models for the diffuse $\gamma$-ray background of the Milky
  Way.  The top panel shows an all-sky map of a simple analytic
  background model given in [8], while the middle and
  bottom panels show background models both for the `conventional' and
  `optimized' GALPROP models of ref.~\cite{Strong2004}, where the
  optimized model has been tuned to approximately reproduce the
  observed EGRET spectrum. For the maps, the analytic model and the
  conventional GALPROP model have been normalized to their mean
  background over the whole sky. The optimized model is 
  normalized to the mean value of the conventional model, to
  illustrate its slightly elevated overall background level.
  \label{FigBackgrounds}}
\end{figure}

First, let us examine the constant background case. Then the $S/N$ of the
MainSm component can be written as
\begin{equation}
(S/N)_{\rm MainSm} =  f_{\rm MainSm} 
\left[ \frac{\tau\, A_{\rm eff} }{b_\gamma}\right]^{1/2} \frac{F}{\theta_h} \, ,
\end{equation}
where we find $f_{\rm MainSm}\simeq 0.498$ from our simulation.

For a variable background, we consider three models, a simple analytic
parameterization given by [8], of the form
\begin{equation}
b_\gamma(l,b) = 0.5+ \left\{
\begin{array}{ll}
\frac{85.5}{\sqrt{[1+(\frac{l}{35})^2][1+(\frac{b}{1.1+0.022|l|})^2]}} &
\mbox{for $|l|\ge 30^{\rm o}$}\\
\frac{85.5}{\sqrt{[1+(\frac{l}{35})^2][1+(\frac{b}{1.8})^2]}} &
\mbox{for $|l| <  30^{\rm o}$}
\end{array}
\right.
\label{EqnBackgrnd}
\end{equation}
where $-180^{\rm o} < l \le 180^{\rm o}$ and $-90^{\rm o} < b \le 90^{\rm o}$
are Galactic coordinates, and two $\gamma$-ray maps for the background
calculated with the GALPROP code for propagation and interaction of cosmic
rays in the Galaxy.  The two GALPROP maps are the `conventional' and
`optimized' models for Galactic $\gamma$-ray emission, as described by Strong
et al.~[31].  We show all-sky maps of these three background
models in Figure~\ref{FigBackgrounds}.

The optimal filter will automatically down-weight the signal where the
background is high, so in principle the $S/N$ can be directly calculated with
these background maps over the whole sky. For this perhaps optimistic
approach, we can again write the $S/N$ as
\begin{equation}
(S/N)_{\rm MainSm} =  f_{\rm MainSm} 
\left[ \frac{\tau\, A_{\rm eff} }{\left<b_\gamma \right>}\right]^{1/2} \frac{F}{\theta_h} \, ,
\end{equation}
where now $\left<b_\gamma\right>$ is the background averaged over the whole sky. We
obtain $f_{\rm MainSm}=0.183$ for the background model of equation
(\ref{EqnBackgrnd}), $f_{\rm MainSm}=0.205$ for the conventional GALPROP map,
and $f_{\rm MainSm}=0.198$ for the optimized GALPROP map. The differences
between the background models are hence quite small, and the $S/N$ coefficient
reduces approximately by a factor $\sim 2.5 - 2.8$ compared with the more
favorable constant background. If the background is averaged only over the
high latitude sky (say for $b>10^{\rm o}$) to account for the fact that most
subhalos reside there, the $S/N$ prefactor reduces further by a factor $1.3$
for the analytic model, and by 1.6 for the GALPROP maps.
  
Note that the $S/N$ of subhalos will in general also change for a variable
background, depending on their position; the majority will slightly increase
their relative $S/N$ as they are located at positions where the background
lies below the sky-averaged background, while for the others the relative
$S/N$ will decrease. This effectively broadens the $S/N$ distribution of
subhalos relative to the $S/N$ of the main halo.

However, the above approach of integrating over the whole sky is probably too
optimistic in reality, as the background is not known very accurately close to
the Galactic center and close to the disk. A more conservative approach may
therefore be to cut out this region entirely, for example by excluding the
region 5 degrees above and below the Galactic plane. This $|b|<5^{\rm o}$ cut
disregards a bit less than 10\% of the sky, but since the main halo's emission
is highly concentrated towards the centre, the impact on the $S/N$ will be
larger. We find that the coefficient $f_{\rm MainSm}$ drops by a factor $2.04$
in this case when the analytic background model given in equation
(\ref{EqnBackgrnd}) is used. If the cut is enlarged to exclude $|b|<10^{\rm
  o}$ instead, the drop is instead a factor of $3.00$. For the GALPROP models,
the corresponding changes are slightly smaller; for the $|b|<5^{\rm o}$ cut,
they evaluate to factors of 1.53 and 1.44 for the conventional and optimized
models, respectively, while for the $|b|<10^{\rm o}$ cut the numbers are 2.21
and 2.10, respectively.

\subsubsection*{Component ``MainUn'', unresolved subhalos}

As we have seen in the Letter, an extrapolation of the simulated
substructure mass spectrum down to $10^{-6}\,{\rm M}_\odot$ suggests
that the cumulative luminosity of all substructures in fact dominates
the total emission of the Milky Way as seen by a distant observer,
being larger by a factor $\sim 232$ than the main halo's diffuse
luminosity within a radius of $246\,{\rm kpc}$. However, most of these
substructures are much further from the Sun than the Galactic
centre, so the total substructure flux received on Earth is
only a factor 1.9 larger than that from the smooth main halo.

We expect this total substructure flux to be nearly uniformly distributed over
the sky, which will make it hard to detect in practice. To obtain a model for
this distribution, we first construct a map of the flux of all resolved
substructures, each of them represented by the  sum of their SubSm and
SubSub luminosities, the latter estimated as described in the Letter. In total,
the SubSub luminosity assigned to the subhalos corresponds to 4.9 times the
total SubSm luminosity, most of it in objects at $R>20\,{\rm kpc}$. We then
still miss the luminosity from independent subhalos below our mass resolution
limit, as well as from their (sub-)substructure, which is the component
MainUn. Based on our extrapolation of the total substructure luminosity within
$r_{200}$ as a function of mass resolution, we know that this raises the total
luminosity contributed by substructures (SubSm + SubSub + MainUn) to 232 times
the main halo's luminosity (MainSm), with a radial luminosity distribution
that is self-similar to that of all resolved subhalos.

If we assume that the halo is spherically symmetric, we can use an
analytic fit to the radial luminosity distribution of resolved
substructures to calculate the surface brightness of the MainUn
component over the sky. We then find that it contributes 85.7 times as
much flux in total as the resolved subhalos and their
(sub-)substructures (SubSm+SubSub), and nearly 1.9 times as much flux
as the main halo (MainSm), but it has a maximum contrast of only 1.54
between Galactic center and anticenter.  We caution however that small
distortions in the halo shape in its outer parts could easily
overwhelm this small contrast, so that one should not expect this gradient
to be detectable in practice. In fact it is probably not even
guaranteed that the center will be the brightest direction. Another
plausible model for distributing the MainUn component is to assume
that it is strictly proportional to the sum of the SubSm and SubSub
components, but very heavily smoothed due to the huge number of
unresolved substructures. For a smoothing of 30 degrees, the contrast
between Galactic center and anticenter varies with observer position
and is typically a factor 1.3 - 2.0, but there are fluctuations in
other parts of the sky that are of similar amplitude.

In practice, it will be impossible to separate the MainUn component
from DM annhihilation in more distant objects. We therefore exclude
this component from our analysis in the Letter (see also below).

\subsubsection*{Ratios of S/N for different components}

As is evident from the above expressions, the observation time and telescope
aperture always drop out in ratios of different $S/N$ values. Likewise, these
ratios are invariant under a multiplicative factor for the background over the
whole sky, provided the background stays sufficiently large that it dominates
everywhere. Also, to the extent that the spectrum of the background is
independent of direction, the spectral energy distribution of the annihilation
radiation is unimportant for $S/N$ ratios.  Our analysis in the Letter is
based on these simplifying assumptions, which allow quite general, and, we
believe, robust statements about the relative detectability of different dark
matter components in the Milky Way.

\begin{figure*}
\rotatebox{90}{%
\begin{minipage}{210mm}
\begin{center} 
\resizebox{10.5cm}{!}{\rotatebox{90}{\includegraphics{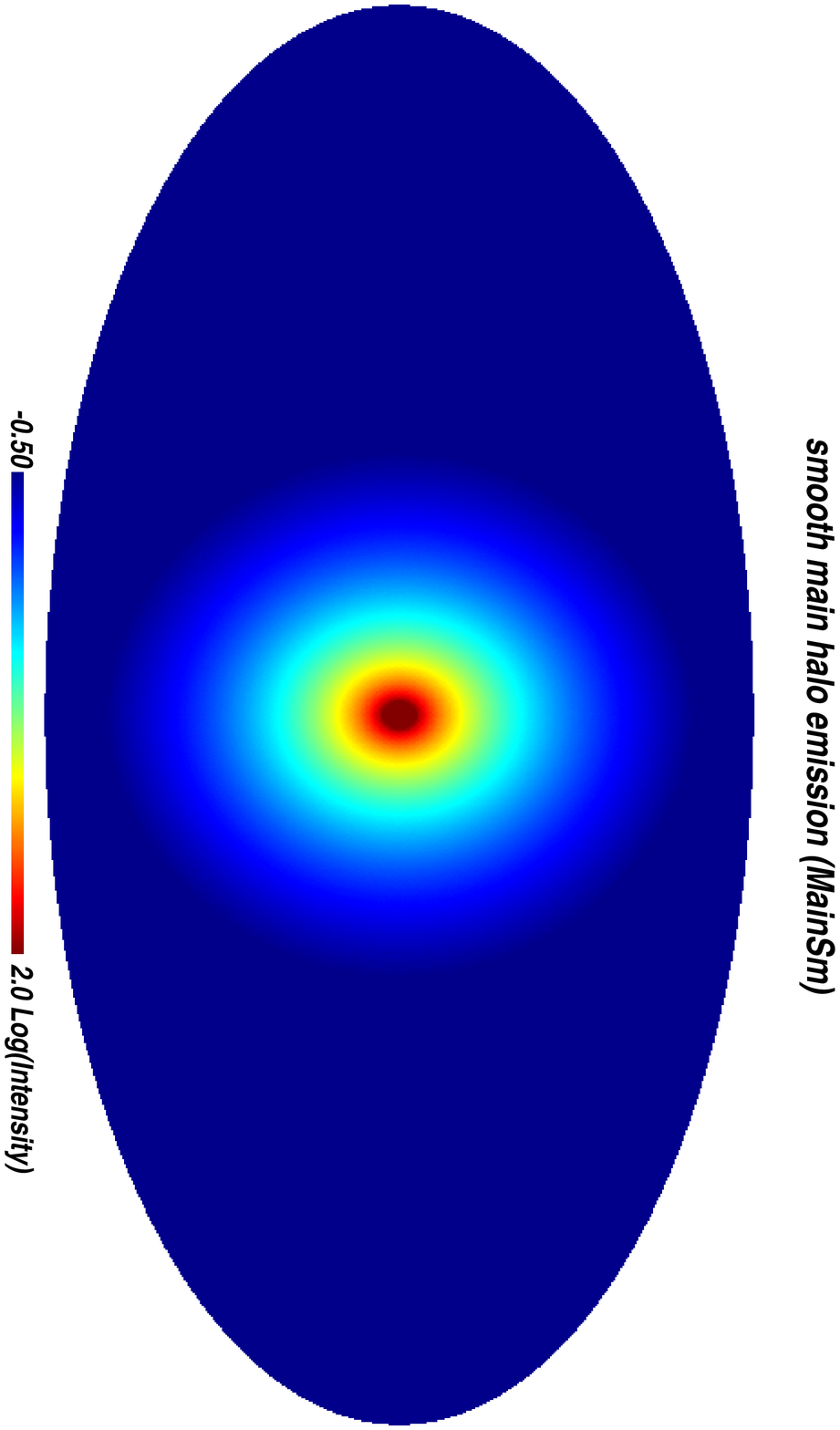}}}%
\resizebox{10.5cm}{!}{\rotatebox{90}{\includegraphics{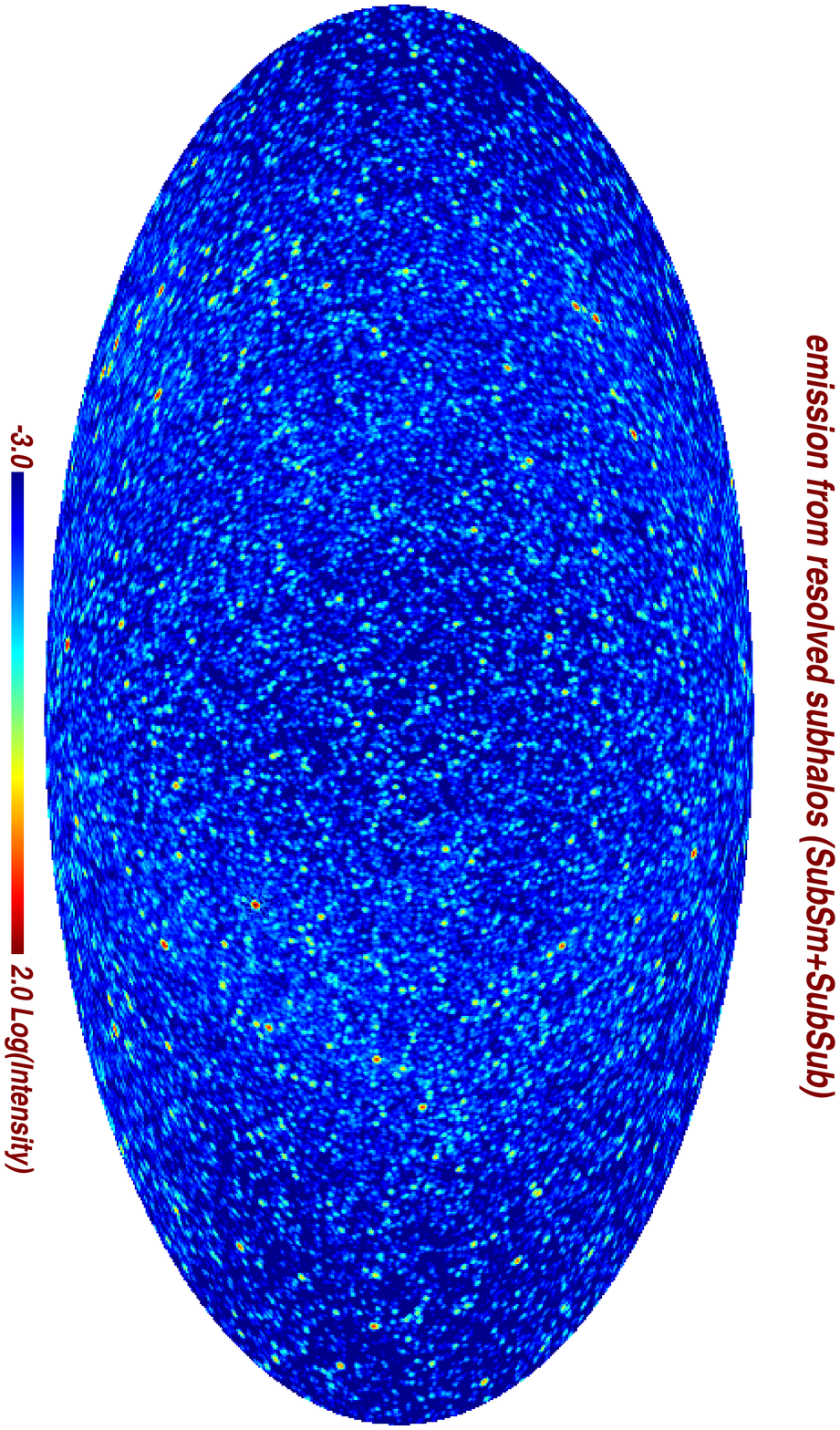}}}\\
\resizebox{10.5cm}{!}{\rotatebox{90}{\includegraphics{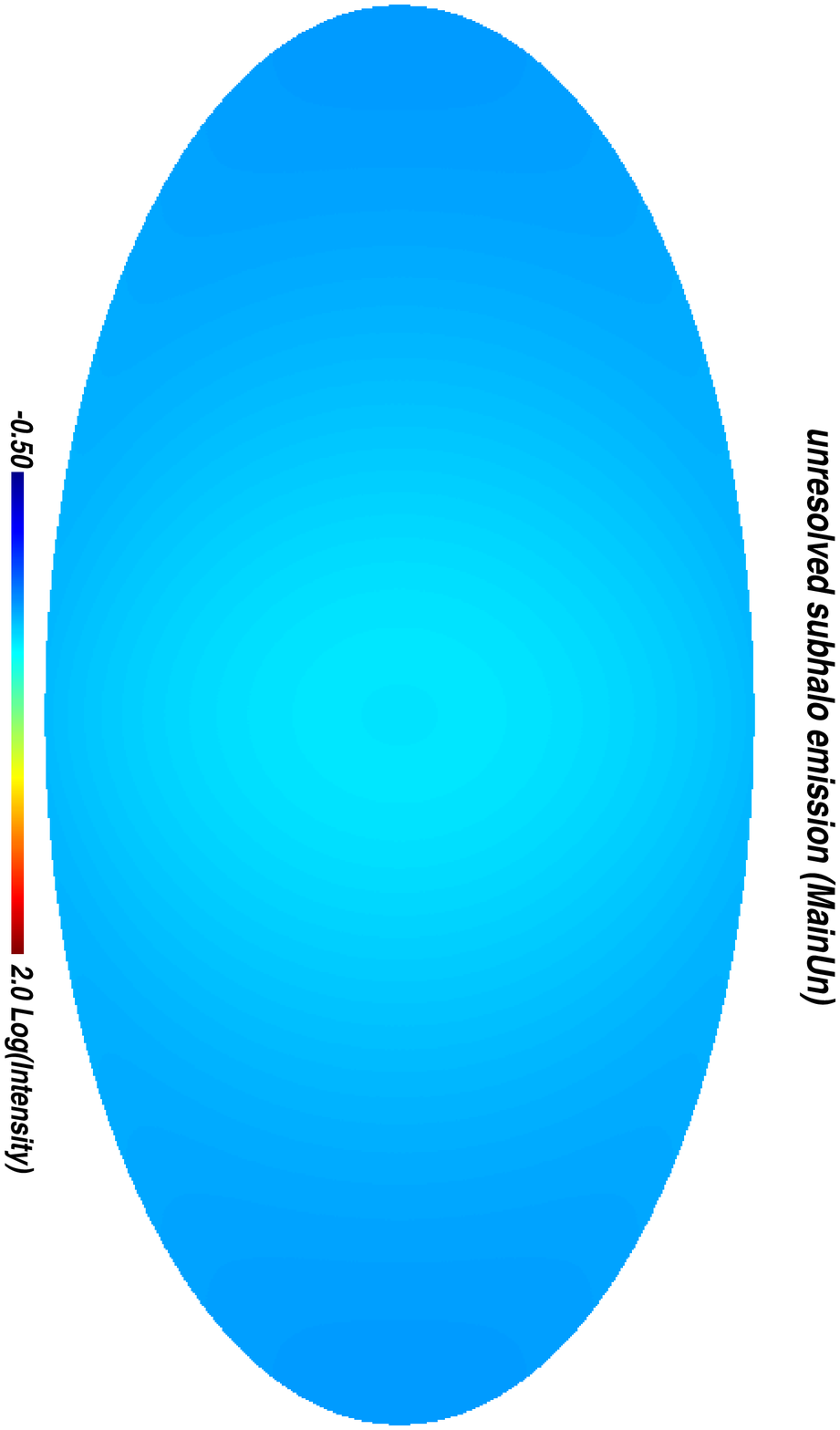}}}%
\resizebox{10.5cm}{!}{\rotatebox{90}{\includegraphics{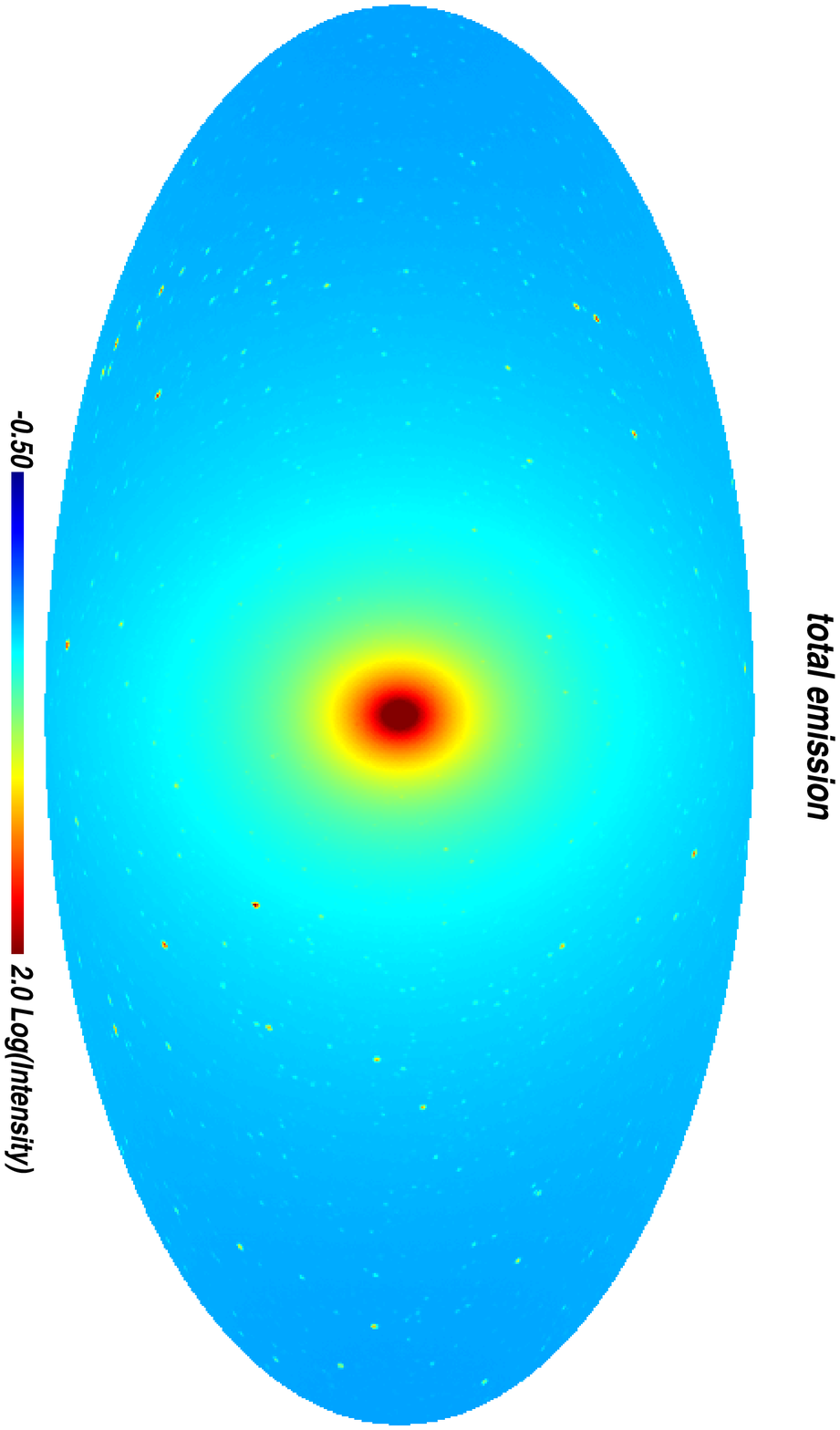}}}\\
\end{center}
\caption{Different emission components. The top left panel shows an
  all-sky map of the main halo's diffuse emission (averaged for
  different observer positions and over azimuth), while the top right
  panel shows the emission from all resolved subhalos, from a random
  position on the Solar circle. The luminosities assigned to each
  subhalo include their contribution for all unresolved
  (sub-)substructure. For simplicity and for better graphical
  reproduction they have been represented as point sources that were
  smoothed with a Gaussian beam of $40\,{\rm arcmin}$.  The bottom
  left panel gives the expected surface brightness from all unresolved
  subhalos down to the free streaming limit, assuming a spherically
  symmetric halo.  This is a very smooth component over the sky that
  dominates the total flux (its integrated flux is nearly 1.9 times
  the integrated flux from the main halo).  Finally, the bottom right
  panel shows the total surface brightness from all components
  together. All maps show the surface brightness in units of the main
  halo's diffuse emission, and use the same mapping to color scale,
  except for the map of the resolved substructures, where the scale
  extends to considerably fainter surface brightness.
  \label{FigEmissionComponents}}
\end{minipage}
}
\end{figure*}

Note again that the near constancy of component MainUn over the sky
will make it very difficult to tell it apart from the truly
extragalactic background (both from annihilating dark matter and from
other sources such as AGN), which has a level that is unknown. The
component MainUn would have to be distinguished from the extragalactic
background by a characteristic spectral signature or by its weak
gradient across the sky. The former is (particle physics)
model-dependent, while the latter will be difficult unless the
Galactic component is dominant. For this reason, we consider that the
main halo smooth component, with its strong gradient and higher mean
surface brightness, is more likely to be detectable even if its total
flux is somewhat lower.

For illustration, Figure~\ref{FigEmissionComponents} shows all-sky maps of the
surface brightness distributions of the different emission components of the
dark matter annihilation signals considered above. The top left panel gives the
azimuthally averaged main halo's diffuse emission. In the top right panel, the
total emission from resolved subhalos is shown. For simplicity and better
visibility in the reproduction of the map, we represent each subhalo as a
point source that is smoothed with a Gaussian beam of $40$ arcmin, assigning
to them the sum of the corresponding SubSm and SubSub emission.  In the bottom
left, we show the rather smooth emission of the component MainUn, assuming a
spherical halo. Finally, the bottom right panel gives the sum of all
components.

\subsection*{Uncertainties}

\subsubsection*{Systematic and random errors}

Our analysis of the signal-to-noise ratios assumed an accurate knowledge of
the background level, so that it is only affected by random errors, in which
case our simple treatment is appropriate. However, in reality background
estimates have a substantial systematic uncertainty, which is perhaps as large
as the difference between the `conventional' and `optimized' GALPROP
models. This introduces systematic uncertainties that require an elaborate
statistical analysis based on a multi-dimensional maximum likelihood
treatment. This is clearly beyond the scope of the present study, but will
certainly become part of the data analysis of the GLAST satellite (see also [3]).

\subsubsection*{Known versus unknown locations}

For simplicity, we have also neglected in our analysis the differences in the
required effective detection threshold between sources at known and unknown
locations. The small dark matter substructures with the highest $S/N$ for
detection should correspond to truly dark subhalos without any stars; hence
their location is initially unknown. As there are of order $5\times 10^5$
resolution elements of $(10\,{\rm arcmin})^2$ on the sky, one would expect
about 640 fluctuations above $3\,\sigma$ just from Gaussian statistics. One
therefore has to detect a dark subhalo at unknown location at least with
$5\,\sigma$ significance before it may qualify as a potential detection, while
for a dwarf satellite at known location, as well as for the main halo, a less
conservative threshold of $3\,\sigma$ may be assumed.

\subsubsection*{Effects of baryons}

In our high-resolution simulations of a Milky Way like halo, we have neglected
the effects of baryons on the growth of dark matter structures. While the dark
matter dominance of the known dwarf satellites suggests that this
approximation should be quite accurate for the dark matter subhalos, the
central dark matter cusp of the main halo is likely modified by baryonic
physics.

Most authors (but not all) argue that baryonic infall due to radiative cooling
and disk formation leads to an adiabatic compression of the halo, which would
enhance the smooth main halo emission relative to the other components we
discuss, thus strengthening our conclusions.  A comparably large variation
comes from the scatter in concentration predicted among different realizations
of halos of Milky Way mass [32].  The halo we analyze here is
approximately a $1\,\sigma$ deviation towards high concentrations in the
expected distribution, while the `Via Lactea'
simulations [13,~33] are 1-$1.5\,\sigma$ deviations in the
opposite sense.  Since baryons are generically expected to increase halo
concentration, it seems appropriate to analyze a halo that lies on the high
side of the concentration distribution.

There have been suggestions in the literature that central dark matter cusps
could perhaps be destroyed by dynamical processes such as stellar bars,
supermassive black hole binaries, or rapid baryonic outflows from explosive
feedback events. We note however that each of these scenarios for cusp removal
is debated and discussed with conflicting numerical results in the literature,
while the effect of adiabatic compression is found robustly in cosmological
simulations with radiative cooling.

We also note that our conclusions do not rely on the dark matter distribution
in the inner few degrees, where additional baryonic effects may play a role,
for example a steepening of the dark matter cusp around the supermassive black
hole at the Galactic centre [34, 35]. As we have shown
above, the $S/N$ for detection of the main halo drops only by a small factor
if this uncertain region around the Galactic centre is excluded.

\subsubsection*{Energy dependence}

An important problem in obtaining conclusive evidence for DM annihilation is
to distinguish the signal from other astrophysical sources. For spatially
resolved sources, the profile of the radiation on the sky can provide such a
discrimination if it can be shown that the signal matches the theoretically
expected dark matter distribution. This should be promising for the smoothly
varying emission from the main halo of the Galaxy, especially if it is
combined with the spectral information obtained from the energy distribution
of the detected photons. The expected spectral shape of the radiation from dark
matter annihilation has a characteristic form with a sharp cut-off near the
mass of the dark matter particle, which is difficult to arrange with other
astrophysical sources. Also, the peak of the annihilation spectrum typically
lies at considerable higher energy in most SUSY models than the peak in the
Galactic diffuse $\gamma$-ray background, which should further help in the
discrimination.

If dark matter substructures are detected by GLAST, we expect them to be
poorly resolved at best, so that their surface brightness profile will be of
little help to identify them as a dark matter signal. Here the background
uncertainty is less of a concern, but the signal must be distinguished from
other astrophysical (point) sources. Again, a multiwavelength approach and the
exploitation of spectral features [36, 37, 38] of the DM annihilation appears
as a highly promising way to convincingly identify the origin of the signal.

\bibliography{si}

\end{document}